\documentclass[showpacs,twocolumn,prb,groupedaddress]{revtex4}
\usepackage{amssymb}
\usepackage{amsmath}
\usepackage{graphicx}

\input{tcilatex}

\begin{document}

\title{Quantum Hall effect of the surface states in topological insulator }
\date{\today }
\author{Shun-Qing Shen}
\affiliation{Department of Physics and Center for Theoretical and Computational Physics,
The University of Hong Kong, Pokfulam Road, Hong Kong, China}

\begin{abstract}
We study the quantum Hall effect in the surface states of topological
insulator in the presence of a perpendicular magnetic field in the framework
of edge states. Motion of Dirac fermions will form descrete Landau levels,
among which a fully saturated zero mode will have different behaviors near
the boundary according to the sign of the effective mass for Dirac fermions.
The Hall conductance is quantized to be $ne^{2}/h$ ($n$ is an integer) for a
positive mass, $(n+1)e^{2}/h$ for a negative mass, and $(n+1/2)e^{2}/h$ for
massless fermions. In topological insulator the massive term $m_{eff}$ to
the Dirac fermions can be the Zeeman coupling in a magnetic field or be
induced by the finite-size effect in an ultrathin film. For example the
g-factor of Bi$_{2}$Se$_{3}$ is positive and give rise to a positive mass
term for Dirac fermions. We address experimental realization of the quantum
Hall effect in topological insulators.
\end{abstract}

\maketitle

Three-dimensional topological insulators possess metallic surface states in
a band gap which reside near the system surface as a consequence of strong
spin-orbit coupling.\cite{Fu-07prl} Recent experiments and the first
principles calculations have verified the existence of this novel type of
topological states in the materials such as Bi$_{1-x}$Se$_{x}$, Bi$_{2}$Se$%
_{3}$, and Bi$_{2}$Te$_{3}$.\cite%
{Hsieh08Nature,Xia-09NP,Zhang-09NP,Chen-09xxx} The surface states have
linear dispersion in the momenta and are modeled as a two-dimensional free
Dirac fermion gas, which is analogous to a single layer graphene. The Hall
conductance of two-dimensional Dirac fermions is well defined only when the
mass $m_{eff}$ of the fermions is nonvanishing, $\sigma _{H}=sgn(m_{eff})%
\frac{e^{2}}{2h}$.\cite{Redlich-84prd} In the presence of a perpendicular
and uniform magnetic field, it was predicted that the Hall conductance of
massless Dirac fermions is quantized as a half of odd integer $(n+1/2)$
timing the quanta $e^{2}/h$ ($n$ is an integer).\cite{Jackiw-84prd} The
charge carriers in a single layer graphene are regarded as massless Dirac
fermions, and the measured quantum Hall conductance is $4(n+1/2)e^{2}/h$
where the factor 4 originates from the spin and valley degeneracy.\cite%
{Zhangyb-05Nature,Novoselov-05Nature,Gusynin-05prl} However, strictly
speaking, the measured quantity in graphene is $2e^{2}/h$, not $e^{2}/2h$
directly. The newly discovered topological insulator Bi$_{2}$Se$_{3}$ and Bi$%
_{2}$Te$_{3}$ have been identified to have a single Dirac cone,\cite%
{Zhang-09NP,Xia-09NP,Chen-09xxx} which may provide an ideal platform to
study a half of the quanta $e^{2}/h$ for the Hall conductance. On the other
hand, the conventional edge state picture for quantum Hall effect \cite%
{Halperin-82prb, MacDonald-84prb} have established an explicit relation
between the Chern number or the quantized Hall conductance and the number of
the edge states.\cite{Hatsugai-93prl} A half-quantized Hall conductance are
challenging the validity or applicability of the edge state theory if the
quantum Hall effect exist in the surface states of topological insulators.

In this paper we study the quantum Hall effect for a two-dimensional Dirac
fermions in the presence of a magnetic field in the edge state theory. The
Landau level formation will lead to the quantization of the quantum Hall
conductance. It is found that a fully saturated zero mode of the Landau
level appears, and its edge effect near the boundary determines its
contribution to the quantum Hall conductance. The sign of the effective mass
of Dirac fermions plays a decisive role. For Dirac fermions of positive
mass, the Hall conductance is $ne^{2}/h$ as a conventional two-dimensional
electron gas, for Dirac fermions of negative mass, the Hall conductance is $%
(n+1)e^{2}/h$, and for massless Dirac fermions the Hall conductance is $%
(n+1/2)e^{2}/h$. These results can be applied to the surface states of
topological insulator and an ultra-thin film of topological insulator. Since
the surface states in topological insulator has a spin texture structure,
which is distinct from the Dirac cone in graphene, the application of
magnetic field breaks the time reversal symmetry and the Dirac fermions in
the surface states acquire a massive term, i.e., the Zeeman term. Thus the
sign of the g-factor will determined the value of quantum Hall conductance.
For example the g-factor in Bi$_{2}$Te$_{3}$ is positive and generates a
positive mass. As a result the quantum Hall conductance is $ne^{2}/h$, which
is identical as a conventional two-dimensional electron gas. For an
ultra-thin film, the finite size confinement will mix the surface states at
the top and bottom layers and generate an energy gap for Dirac fermions.
Since the system is still invariant under time reversal symmetry, one set of
fermions possesses positive mass and another set possesses negative mass.%
\cite{Lu-09xxx} As a result the Hall conductance can be $2(n+1/2)e^{2}/h$ if
the energy gap is larger than the Zeeman splitting.

We start with a 2+1 massive Dirac Hamiltonian%
\begin{equation}
H=v_{F}\hbar (k_{x}\sigma _{x}+k_{y}\sigma _{y})+m_{eff}v_{F}^{2}\sigma _{z}
\end{equation}%
where $v_{eff}$ is the effective speed of light and $m_{eff}$ is the
effective mass. $\sigma _{x,y,z}$ are the Pauli matrices. To calculate the
Hall conductance the Hamiltonian can be written in the form, $H=d(k)\cdot
\sigma $ with the vector $\mathbf{d}(k)=(v_{F}\hbar k_{x},v_{F}\hbar
k_{y},m_{eff}v_{F}^{2})$. The Hall conductance of the Dirac fermions is only
well defined in a massive case, 
\begin{equation}
\sigma _{H}=-\frac{e^{2}}{2\hbar }\sum_{k}\frac{\mathbf{d}\cdot (\partial
_{k_{x}}\mathbf{d}\times \partial _{k_{y}}\mathbf{d})}{\left\vert
d(k)\right\vert ^{3}}=sgn(m_{eff})\frac{e^{2}}{2h}
\end{equation}%
when the Fermi energy is located between the band gap.\cite%
{Redlich-84prd,Qi-08prb,Zhou-06prb} For a non-interacting system, it is
impossible to have a half-quantized Hall conductance in unit of $e^{2}/h$.
However, the sign change of the mass $m_{eff}$ will lead a jump of Hall
conductance, i.e., $\Delta \sigma _{H}=\frac{e^{2}}{2h}$.

\FRAME{ftbpFU}{3.3425in}{2.0998in}{0pt}{\Qcb{(a). Geometry of the sample
subjected to a perpendicular magnetic field B. The edge state spectrum of
the Landau levels (b) for Dirac fermions with negative mass, (c) for a
massless Dirac fermions, and (d) for Dirac fermions with positive mass. }}{}{%
edgestates.png}{\special{language "Scientific Word";type
"GRAPHIC";maintain-aspect-ratio TRUE;display "USEDEF";valid_file "F";width
3.3425in;height 2.0998in;depth 0pt;original-width 3.8467in;original-height
2.405in;cropleft "0";croptop "1";cropright "1";cropbottom "0";filename
'edgestates.png';file-properties "XNPEU";}}

Now we come to study the formation of the Landau levels with a finite
boundary following the theories of edge states by Halperin \cite%
{Halperin-82prb} and MacDonald and Strada\cite{MacDonald-84prb}. We first
consider a geometry of strip with a width $L_{y}$ and thickness $H$, which
are much larger than the magnetic length $l_{B}$ and the spatial
distribution $\xi $\ of the surface states. Assume the magnetic field $B$
(alone the z-axis) is perpendicular to the slab as shown in Fig. 1(a). We
first focus on the top plane. The periodic boundary condition is taken along
the x-axis, and the open boundary condition along the y axis. In this way
the wave number $k_{x}$ is still a good quantum number, and $k_{y}$ is
substituted by $-i\partial _{y}$. We take the Landau gauge for the vector
potential, $A_{x}=+By$ and $A_{y}=0$. Thus we define $a(y_{0})=\frac{l_{B}}{%
\sqrt{2}}[\partial _{y}+l_{B}^{-2}(y-y_{0})]$ where the magnetic length $%
l_{B}=\sqrt{\hbar /eB}$ and $y_{0}=-l_{B}^{2}k_{x}$ assuming $eB>0$. The
operators $a$ and $a^{\dag }$ satisfy the commutation relation, $%
[a(y_{0}),a^{\dag }(y_{0})]=1$. For simplicity, we introduce a dimensionless
parameters $\delta =m_{eff}v_{F}^{2}l_{B}/(\sqrt{2}v_{F}\hbar )$. In this
way, we have a dimensionless Schrodinger equation, 
\begin{equation}
\left( 
\begin{array}{cc}
\delta  & a \\ 
a^{\dag } & -\delta 
\end{array}%
\right) \left( 
\begin{array}{c}
\varphi _{1} \\ 
\varphi _{2}%
\end{array}%
\right) =\frac{E}{v_{F}\sqrt{2e\hbar /B}}\left( 
\begin{array}{c}
\varphi _{1} \\ 
\varphi _{2}%
\end{array}%
\right) .  \label{a-eq}
\end{equation}%
The allowed values for $y_{0}$ are separated by $\delta y_{0}=2\pi
l_{B}^{2}/L_{x}$ with a periodic boundary condition with length $L_{x}$ and
are limited within $0<y_{0}<L_{y}$. The solution is a function of the good
quantum number $k_{x}$ or $y_{0}=-l_{B}^{2}k_{x}$. When $y_{0}$ is far away
from two edges of $y=0$ and $y=L$, the two components $\varphi _{1}$ and $%
\varphi _{2}$ will vanish at the two boundaries. In this case, the energy
eigenstates are 
\begin{equation}
\left\vert n,\alpha \right\rangle =\left( 
\begin{array}{c}
\sin \theta _{n,\alpha }\left\vert n-1\right\rangle  \\ 
\cos \theta _{n,\alpha }\left\vert n\right\rangle 
\end{array}%
\right) 
\end{equation}%
where $\tan \theta _{n,\alpha }=\frac{\sqrt{n}}{\alpha \sqrt{n+\delta ^{2}}%
-\delta },$ $\alpha =\pm 1$, $n$ is a positive integer, $\left\vert
n\right\rangle =\frac{1}{(n!)^{1/2}}(a^{\dag }(y_{0}))^{n}\left\vert
0\right\rangle $ the solution for a simple harmonic oscillator and $%
a(y_{0})\left\vert 0\right\rangle =0$.\cite{Shen-04prl} The Landau energy is
given by%
\begin{equation}
E_{n,\alpha }=\alpha v_{F}\sqrt{2e\hbar B\left( n+\delta ^{2}\right) }
\end{equation}%
for a positive integer $n$, which are highly degenerate.

It should be emphasized that the zero mode $E_{0}=-v_{F}\sqrt{2e\hbar B}%
\delta $ for $n=0$ and the eigenstate is fully saturated, $\left\vert
0,0\right\rangle =\left( 
\begin{array}{c}
0 \\ 
\left\vert 0\right\rangle 
\end{array}%
\right) $. The number of the allowed values of $y_{0},$\thinspace $%
N_{L}=L_{y}/\delta y_{0}=2\pi L_{x}L_{y}/l_{B}^{2},$ is the degeneracy of
the Landau levels. The energy expressions yield an energy gap $\Delta
E=\left\vert E_{n=\pm 1}\right\vert -E_{0}$ between the zero mode and the
states of $n=\pm 1$. For $\delta =0$ the energy gap is about $\Delta
E\approx 800$K for Bi$_{2}$Se$_{3}$ at $B=10$T, which makes it possible that
the quantum Hall effect can be measured even at room temperature just as in
single layer graphene.\cite{Novoselov-07SCI}

The boundary effect will remove the degeneracy of the Landau level when $%
y_{0}$ is near the boundary. It was known that we cannot simply take a rigid
wall condition such that the two components $\varphi _{1}$ and $\varphi _{2}$
will vanish simultaneously at the boundary. No solution in Eq.(\ref{a-eq})
will satisfy this type of boundary conditions. In the graphene it is a
common practice to calculate the boundary effect numerically in the tight
binding approximation.\cite{Abanin-07ssc} In the present case, since the
surface state has a finite spatial distribution, we do not expect the two
components $\varphi _{1}$ and $\varphi _{2}$ will vanish simultaneously near
the boundary. We first focus on the boundary effect to the zero mode. After
some algebra, we have a expression for the zero energy mode

\begin{equation}
E_{\nu _{0}}=-sgn(\delta )v_{F}\sqrt{2e\hbar B(\nu _{0}+\delta ^{2})}
\end{equation}%
where $\nu _{0}=\left\langle \varphi _{2}\right\vert a^{\dag
}(y_{0})a(y_{0})\left\vert \varphi _{2}\right\rangle /\left\langle \varphi
_{2}|\varphi _{2}\right\rangle .$ The sign of $\delta $ guarantees the
continuity of the spectrum from the bulk to the edge. When $0<<y_{0}<<L_{y}$%
, $\nu _{0}=0$ and $\varphi _{2}(y,y_{0})\propto \left\langle
y|n=0\right\rangle =\frac{1}{\pi ^{1/4}l_{B}^{1/2}}\exp
[-(y-y_{0})^{2}/2l_{B}^{2}].$ If the boundary condition makes $\varphi _{2}=0
$ for $y_{0}=0$ we have a solution for $\varphi _{2}(y,y_{0}=0)\propto
\left\langle y|n=1\right\rangle =\frac{\sqrt{2}}{\pi ^{1/4}l_{B}^{3/2}}y\exp
[-y^{2}/2l_{B}^{2}].$\cite{MacDonald-84prb} In this case $\nu _{0}=1$, but
in this case $\varphi _{1}(y,y_{0})\propto \left\langle y|n=0\right\rangle $%
. The value of $\nu _{0}$ varies from 0 to 1 when $y_{0}$ moves from the
bulk to $y_{0}=0$ or $L_{y}.$ In another limit, if we take $\varphi _{1}=0$
for $y_{0}=0$, the solution $\left\vert 0,0\right\rangle =\left( 
\begin{array}{c}
0 \\ 
\left\vert 0\right\rangle 
\end{array}%
\right) $ and is not distorted by the boundary at all, which is apparently
unphysical. It will be reasonable to assume $0<\nu _{0}<1$ near the
boundaries, which is between these two limits, since the boundaries should
distort the Gaussian-type wave function in the bulk ($0<<y_{0}<<L_{y}$). For
a negative $\delta $, the spectra of the Landau levels goes upward near the
boundary like fermions of positive mass, but for a positive $\delta $ the
spectrum goes like fermions of negative mass. For massless fermions of $%
\delta =0$, the energy spectra near the boundary have two possible
solutions, $E_{\nu _{0}}=+v_{F}\sqrt{2e\hbar B\nu _{0}}$ and $E_{\nu
_{0}}=-v_{F}\sqrt{2e\hbar B\nu _{0}}$. Consider the particle-hole symmetry.
We may say half of the particles in the zero mode have positive energy while
another half of the particles have negative energy. For other states of an
integer $n$, $\nu _{n}=\left\langle \varphi _{2}\right\vert a^{\dag
}(y_{0})a(y_{0})\left\vert \varphi _{2}\right\rangle /\left\langle \varphi
_{2}|\varphi _{2}\right\rangle $ is between $n$ (in the bulk) and $2n+1$
near the edge boundary. In general case, we plot the energy spectra
schematically in Fig. 1(b), (c) and (d) according to the value of $\delta .$

According to the sign of the mass, $m_{eff},$ these energy spectra near the
boundary will lead to three different results for quantum Hall conductance.
For $\delta <0$, the zero mode has the same behaviors of the states of $n>0$
and $\alpha =+1$: the spectrum goes upward near the boundary. It will
generate an edge current. Following MacDonald and Streda,\cite%
{MacDonald-84prb} the edge current is given by 
\begin{eqnarray}
I_{n=0} &=&\frac{1}{\delta y_{0}L_{y}}\int_{0}^{L_{y}}dy_{0}\frac{e}{h}%
l_{B}^{2}\frac{\partial E_{\nu _{0}}(y_{0})}{\partial y_{0}}\Theta (\mu
-E_{\nu _{0}}(y_{0}))  \notag \\
&=&\frac{e}{h}(\mu ^{R}-\mu ^{L}),
\end{eqnarray}%
where $\mu ^{R,L}$ are the potentials at the two sides. This will contribute
one $e^{2}/h$ to the Hall conductance. Thus the Hall conductance should be $%
\sigma _{xy}=(n+1)\frac{e^{2}}{h}$ for $\delta <0$ when other $n$ Landau
levels above the zero mode are filled. For $\delta >0$, the energy spectrum
goes downward near the boundary. The zero mode will not contribute to the
Hall conductance when the Fermi energy is above the the zero mode, $%
E_{f}>E_{\nu _{0}}$, and thus the Hall conductance is $n\frac{e^{2}}{h}$,
which is the quantum Hall conductance for a conventional two-dimensional
electron gas. The massless case of $\delta =0$ is special since the spectra
near the boundary can go either upward or down-ward. We may say half of the
fermions in the zero modes have positive mass, and another half have
negative mass. In this case the zero mode will contribute one half of the
quanta$\frac{e^{2}}{h}$ to the Hall conductance. As a result, the Hall
conductance should be $(n+\frac{1}{2})\frac{e^{2}}{h}$, which is in
agreement with the previous results.\cite{Redlich-84prd} In short, the
quantum Hall conductance for the surface states is determined by the sign of 
$\delta $ or $m_{eff}$,%
\begin{equation}
\sigma _{H}=\left\{ 
\begin{array}{ccc}
(n+1)\frac{e^{2}}{h} & \text{if} & \delta <0 \\ 
(n+\frac{1}{2})\frac{e^{2}}{h} & \text{if} & \delta =0 \\ 
n\frac{e^{2}}{h} & \text{if} & \delta >0%
\end{array}%
\right. .  \label{hall-conductance}
\end{equation}

Now turn to apply these results to the Dirac fermions in the surface states
of topological insulators. The Dirac fermions are massless and the Dirac
point is protected by the time reversal symmetry, and is robust against
impurities or defects. Since the Dirac fermions carries real spin with a
texture structure in the momentum space, a Zeeman splitting will be induced
when the system is subjected to a perpendicular magnetic field, $g_{eff}\mu
_{B}\sigma _{z}$, which is equivalent to the effective mass term $%
m_{eff}=g_{eff}\mu _{B}/v_{F}^{2}$. Thus the sign of the g-factor will
determine the value of quantum Hall conductance as in Eq. (\ref%
{hall-conductance}). For example in Bi$_{2}$Se$_{3}$, numerical estimation
of the value $g_{eff}$ from the tight binding approximation is about $%
g_{eff}=1.8$.\cite{Note-Fang} This illustrates that the Dirac fermions in
the surface states acquires a positive mass in the presence of a
perpendicular magnetic field. From this result, we conclude that the quantum
Hall conductance of the surface states in Bi$_{2}$Se$_{3}$ is $n\frac{e^{2}}{%
h}$ as in two-dimensional electron gas.

Consider the magnetic field normal to the top and bottom surfaces of a
sample as shown in Fig. 1. The surface states of all lateral sides just
experience a in-plane field, in which the main effect of magnetic field is
the Zeeman coupling. We have discussed the quantum Hall conductance in the
top surface. If we assume the magnetic field $B$ is along the z-direction,
normal to the top surface, the electrons in the bottom surface will
equivalently experience a $-B$ field. In this case, the operators $a$ and $%
a^{\dag }$ in the effective model (Eq.(\ref{a-eq})) for the bottom surface
fermions should be re-defined: $a$ operator should be replaced by $a^{\dag }$
and vice versa. The $\delta $-term is attributed to the Zeeman term, it will
also change its sign, $\delta \rightarrow -\delta $. As a result, the zero
energy mode will change to $\left\vert 0,0\right\rangle =\left( 
\begin{array}{c}
\left\vert 0\right\rangle  \\ 
0%
\end{array}%
\right) $ and the energy remains unchanged, $E_{0}=-v_{F}\sqrt{2e\hbar B}%
\delta $. Thus we conclude that the Hall conductance $\sigma _{xy}$ of the
bottom surface is identical to the top surface, and there is no voltage drop
between the top and bottom surface states. Thus the total Hall conductance
should be $2ne^{2}/h$ ($g_{eff}>0$) or $2(n+1)e^{2}/h$\ ($g_{eff}<0$). These
properties will make feasible to measure the quantum Hall effect of the
topological surface states.

The finite size effect \cite{Zhou-08prl} of the surface states will generate
different quantum Hall effect in an ultra-thin film of topological
insulator. When the spatial distribution of the wave function of the surface
states is comparable with the thickness of the thin film, the quantum
tunneling of the wave functions of the top and bottom surface states will
generate an energy gap in the dispersion of Dirac fermions.\cite%
{Lu-09xxx,Linder-09xxx,Liu-09xxx} Since the system is still invariant under
the time reversal symmetry, the two sets of Dirac fermions, as a mixture of
the top and bottom surface states, will acquires positive and negative
masses, $m_{eff}v_{F}^{2}=\tau _{z}\Delta $ with $\tau =\pm 1$,
respectively. The energy gap is estimated to be about $0.13$eV when the
thickness of thin film is 20\r{A}. The effective mass term for Dirac
fermions in an ultra-thin film becomes 
\begin{equation}
\Delta H=(\tau _{z}\Delta +g_{eff}\mu _{B}B)\sigma _{z}.
\end{equation}%
The value of the energy gap $\Delta $ is dependent of the film thickness and
decays exponentially to zero. If the gap is smaller than the Zeeman
splitting, the Hall conductance will be determined by the sign of the
g-factor. However, if the gap is larger than the Zeeman splitting, the
effective mass is positive for one set of fermions and negative for another
set. In this case the quantum Hall conductance for the two sets of Dirac
fermions are $(n+1)\frac{e^{2}}{h}$ and $n\frac{e^{2}}{h},$ respectively. As
a result the total quantum Hall conductance is $2(n+\frac{1}{2})\frac{e^{2}}{%
h}$, which looks like the\ Hall conductance for doubly degenerated and
massless Dirac fermions. Lee \cite{Lee-09xxx} obtained the quantized Hall
conductance $2(n+\frac{1}{2})\frac{e^{2}}{h}$ in the same geometry, which is
consistent with the present result by neglecting the Zeeman splitting.

In conclusion, the Zeeman coupling will be induced as a mass term for Dirac
fermions in the topological surface states in the presence of a magnetic
field. From the point of view of edge states, the sign of the effective
g-factor or mass determines the dispersion behaviors of Landau levels near
the boundary, and further the value of the quantum Hall conductance of the
Dirac fermions. For Bi$_{2}$Se$_{3}$ the calculated g-factor is positive, or
equivalently the Dirac fermions have positive mass, which gives rise to the
quantum Hall conductance $2ne^{2}/h,$as the conventional two-dimensional
electron gas. For an ultra-thin film, the finite-size effect will open an
energy gap for Dirac fermions. If the gap is smaller than the Zeeman
splitting, the Hall conductance is $2ne^{2}/h$ while if it is larger than
Zeeman splitting, the quantum Hall conductance is $2(n+\frac{1}{2})\frac{%
e^{2}}{h}$.

The author would like to thank Zhong Fang for helpful discussions and
numerical calculation of the g-factor in Bi$_{2}$Se$_{3}$. This work was
supported by the Research Grant Council of Hong Kong under Grant No.: HKU
7037/08P, and HKU 10/CRF/08.

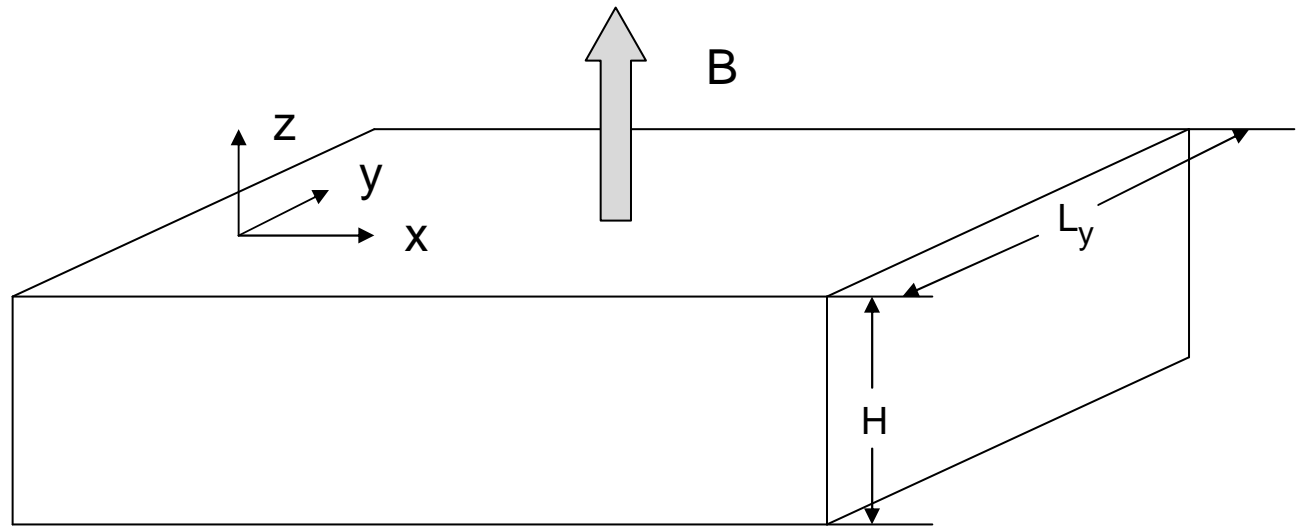
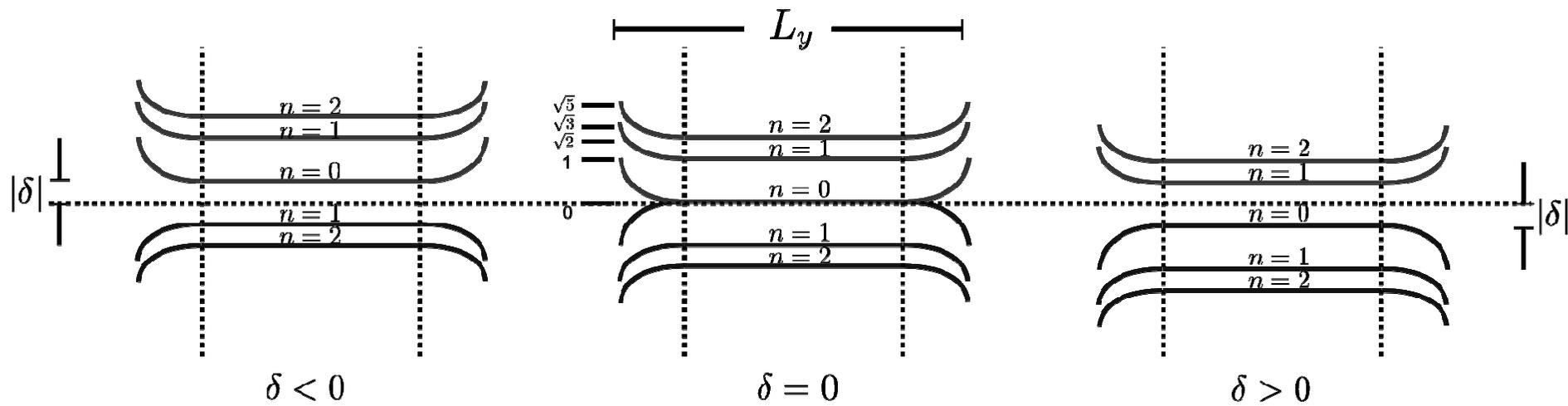
\end{document}